\documentclass[aps,pra,twocolumn]{revtex4-2}

\usepackage{dcolumn}
\usepackage[colorlinks=true]{hyperref}
\usepackage{siunitx}
\usepackage{graphicx}
\usepackage{amsmath}
\usepackage{amssymb}
\usepackage{multirow}%
\usepackage{xcolor}%
\usepackage[version=4]{mhchem}
\usepackage{braket}

\begin{document}

\title{Quantum memory on a nanophotonic silicon chip}

\author{Stephan Rinner}
\author{Jonas Schmitt}
\author{Kilian Sandholzer}
\author{Andreas Reiserer}  \email{andreas.reiserer@tum.de}

\affiliation{Technical University of Munich, TUM School of Natural Sciences, Physics Department and Munich Center for Quantum Science and Technology (MCQST), James-Franck-Stra{\ss}e 1, 85748 Garching, Germany}
\affiliation{TUM Center for Quantum Engineering (ZQE), Am Coulombwall 3A, 85748 Garching, Germany}
\affiliation{Max Planck Institute of Quantum Optics, Quantum Networks Group, Hans-Kopfermann-Stra{\ss}e 1, 85748 Garching, Germany}
\date{\today}

\begin{abstract}

Integrated photonic circuits offer great promise for quantum technologies. However, due to the rapid propagation of light, many envisioned applications require efficient on-chip quantum memories with a programmable delay, compact footprint, and high fidelity. Implementing this based on standard semiconductor processing technology is an outstanding challenge. Here, we realize such memories using erbium-doped silicon waveguides, fabricated as part of a multi-wafer project by a nanophotonic foundry. We demonstrate light storage with a \SI{44.2 \pm 0.9}{\mega\hertz} bandwidth and a programmable delay exceeding \SI{1}{\micro \second} in a device with a footprint of only \SI{1.5e-2}{\milli \meter \squared}, outperforming on-chip delay lines by many orders of magnitude. The phase of the read-out light field is preserved with a visibility of \SI{91.3 \pm 3.0}{\percent}. The efficiency of \SI{1.89 \pm 0.28 e-8}{} can be improved in future devices through resonator enhancement and higher dopant concentrations. With this, the demonstrated approach will pave the way towards applications in photonic quantum computing based on scalable silicon processing technology. 

\end{abstract}


\maketitle

\section{Introduction}\label{Sec:Introduction}

Quantum computers may dramatically outperform their classical counterparts in many applications, including database search, optimization problems, machine learning, solution of linear systems of equations, and simulation of complex quantum systems~\cite{alexeev_quantum_2021}. To unfold their full potential, the size and fidelities achieved in current noisy intermediate-scale quantum (NISQ)~\cite{preskill_quantum_2018} computers need to be increased, allowing the operation of millions of qubits and enabling efficient quantum error correction~\cite{breuckmann_quantum_2021}. In this context, integrated photonic quantum technologies offer great promise~\cite{pelucchi_potential_2021, aharonovich_programmable_2026}: Photonic qubits can be manipulated and detected with high fidelity~\cite{maring_versatile_2024} and free of cross-talk; corresponding devices can be mass-fabricated by standard semiconductor nanofabrication technology~\cite{alexander_manufacturable_2025, larsen_integrated_2025}. In addition, using only linear operations and measurements, small-scale entangled states can be generated and grown into large clusters, enabling universal computation~\cite{rudolph_why_2017, slussarenko_photonic_2019, takeda_toward_2019, romero_photonic_2025}.

However, optical qubits also have a drawback: because photons propagate rapidly, they cannot be kept for extended periods. Thus, preserving quantum information over many clock cycles of a quantum computer requires continuously generating and measuring photonic qubits, resulting in significant overhead and making the system prone to errors~\cite{rudolph_why_2017, takeda_toward_2019, romero_photonic_2025}. This difficulty may be mitigated by integrating quantum memories into photonic chips, enabling more efficient computing architectures and distributed quantum information processing systems, including quantum networks~\cite{wehner_quantum_2018} and repeaters~\cite{loock_extending_2020}.

Such quantum memories should be compact and enable efficient storage of photonic quantum bits for extended periods, as well as retrieval on demand or after a programmable delay. In addition, the memory bandwidth should match the $\sim \si{\giga\hertz}$ clock cycle of photonic quantum computers, and that of efficient solid-state sources of single photons~\cite{holewa_solid-state_2025} or other non-classical states~\cite{takeda_toward_2019, larsen_integrated_2025}. Finally, the devices should operate in the telecommunications frequency band, where losses in optical fibers are minimal and robust lasers and photonic components are readily available.

Previous experiments investigated fiber-optical~\cite{takeda_toward_2019, madsen_quantum_2022, aghaee_rad_scaling_2025} or on-chip silicon photonic delay lines~\cite{hong_multimode-enabled_2025}; however, this precludes programmable delay, and the comparably large dimensions of such devices hinder dense integration. Furthermore, in on-chip devices, the waveguide loss, around \SI{10}{\decibel \per \meter} in optimized silicon~\cite{aalto_open-access_2019} and below \SI{1}{\decibel \per \meter} in optimized SiN waveguides~\cite{liu_high-yield_2021}, restricts the maximum achievable delay to \SI{\sim 20}{\nano\second} at efficiencies $>50\,\%$. 

Therefore, integrated photonic quantum memories based on solid-state photon emitters have been explored as an alternative~\cite{zhou_photonic_2023}. Among others, the atomic-frequency comb (AFC) protocol~\cite{afzelius_multimode_2009} offers programmable delay and high efficiency while supporting multiple temporal and/or spectral modes. AFC memories based on rare-earth dopants have been demonstrated in different bulk crystals, including Yttrium Orthosilicate \cite{tiranov_storage_2015, ortu_multimode_2022, sanchez_mejia_broadband_2025, feldmann_cavity-enhanced_2025, lv_minute-scale_2025, jiang_quantum_2023, alqedra_stark_2024, craiciu_multifunctional_2021, stuart_initialization_2021}, Yttrium Orthovanadate~\cite{zhou_quantum_2015}, and others \cite{bonarota_highly_2011, davidson_improved_2020, falamarzi_askarani_long-lived_2021}. Efficiencies exceeding \SI{50}{\percent} have been reached using a cavity-enhanced AFC \cite{duranti_efficient_2024}. Long-term memories based on spin-wave transfer~\cite{afzelius_demonstration_2010} have reached storage times exceeding \SI{1}{\hour}~\cite{ma_one-hour_2021}.

In addition to devices using bulk crystals, integrated photonic memories~\cite{zhou_photonic_2023} have been realized with nanophotonic waveguides in Lithium Niobate~\cite{sinclair_spectral_2014, dutta_atomic_2023}, Yttrium Orthovanadate~\cite{zhong_nanophotonic_2017}, and Yttrium Orthosilicate~\cite{rakonjac_storage_2022, liu_millisecond_2025}. In photonic resonators, efficiencies $>\SI{80}{\percent}$ have been achieved~\cite{meng_efficient_2026}. Storage of light in the telecommunications C-band, where losses in optical fibers are minimal~\cite{holewa_solid-state_2025}, is enabled by erbium-doped devices, including glass fibers~\cite{wei_quantum_2024, grimau_puigibert_entanglement_2020}, Lithium Niobate~\cite{zhang_telecom-bandintegrated_2023, falamarzi_askarani_storage_2019} and Yttrium Orthosilicate waveguides~\cite{craiciu_nanophotonic_2019, liu_-demand_2022}. However, these materials are difficult to integrate into established semiconductor manufacturing procedures. Thus, realizing a functional quantum memory in a foundry-compatible photonic circuit has been an open challenge.

Here, we implement such a device and demonstrate a fully integrated AFC memory in an erbium-doped silicon (Er:Si) waveguide~\cite{gritsch_narrow_2022, rinner_erbium_2023}, commercially fabricated by a nanophotonic foundry in a multi-project wafer run ~\cite{aalto_open-access_2019}. We demonstrate the storage and retrieval of faint light pulses with up to \SI{44.2 \pm 0.9}{\mega\hertz} bandwidth and verify memory coherence via interference measurements. We achieve adjustable storage times exceeding \SI{1}{\micro \second} in a waveguide with a footprint as low as \SI{1.5e-2}{\milli \meter \squared}. For the same storage duration, compact silicon delay lines would require an area more than four orders of magnitude larger, and would exhibit an efficiency around $\SI{e-400}{}$~\cite{hong_multimode-enabled_2025}, limited by the exponential propagation losses. In contrast, our memory achieves \SI{1.89 \pm 0.28 e-8}{}, and further improvements are expected upon future optimization. With this, our approach opens the door to synchronising probabilistic on-chip photon sources~\cite{rudolph_why_2017, slussarenko_photonic_2019} and to photonic quantum computers with integrated memories that can be mass-fabricated.
 
\section{Experiment}  \label{sec:experiment}

\begin{figure}[h!]
\centering
\includegraphics[width=0.5\textwidth]{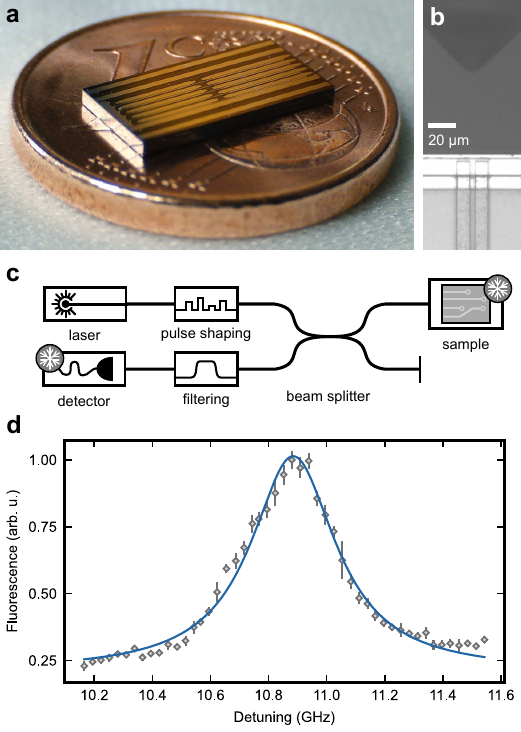}
\caption{\textbf{Setup and device characterization. a,} Photograph of the nanophotonic chip hosting the quantum memory on top of a 1-Euro-cent coin for a size comparison. \textbf{b,} Optical microscope image of the lensed fiber tip (top) used to edge-couple light into one of the waveguides on the silicon nanophotonic chip (bottom). \textbf{c,} Schematic of the fiber-based measurement setup. The light of a frequency-stabilized laser (top left) is switched and frequency-shifted by optical modulators. Using a fiber-based beam-splitter, part of the light is directed to the sample, which is mounted in a closed-cycle cryostat. The light emitted from the sample is guided to a single-photon detection system via a fiber-based temporal and spectral filtering system.  \textbf{d,} Pulsed resonant fluorescence spectrum of the Er ensemble emission at \SI{0.75}{\tesla}. The sample is excited with laser pulses of \SI{15}{\micro \second} duration and the fluorescence is measured within \SI{265}{\micro\second} after the pulses (grey data). A Lorentzian fit (blue) yields a linewidth of \SI{351 \pm 16}{\mega\hertz}. The detuning is defined as the frequency difference with respect to the emission wavelength of Erbium in site A at \SI{0}{\tesla} \SI{1537.7629 \pm 0.00009}{\nano\meter}. Error bars denote one standard deviation after averaging \SI{1e3}{} measurements.} \label{fig1:setup_and_inhom}
\end{figure}

To realize an on-chip quantum memory, we use a commercial silicon-on-insulator (SOI) device fabricated by the VTT foundry~\cite{aalto_open-access_2019}, as shown in Fig. \ref{fig1:setup_and_inhom}a, which hosts \SI{6}{\milli \meter} long and \SI{2.5}{\micro \meter} wide, straight rib waveguides, made by a \SI{1.2}{\micro\meter} deep etch into the \SI{3}{\micro\meter} thick device layer. We integrate erbium dopants into site A~\cite{gritsch_narrow_2022, holzapfel_characterization_2025} by implantation and annealing, with a simulated peak concentration of \SI{1e15}{\per\centi\meter\cubed} (see Methods). This comparably low concentration largely avoids laser-induced spectral diffusion and thus enables long storage times~\cite{fruh_spectral_2026}; however, it compromises the device efficiency, as discussed later. The footprint of the waveguide is $\approx\SI{1.5e-2}{\milli \meter \squared}$; it contains around \num{7.5e6} Er ions from which we expect less than one percent to be located at the studied site A~\cite{gritsch_narrow_2022}.

Except where stated otherwise, the sample is mounted in a closed-cycle $^4\text{He}$ cryostat at $\lesssim\SI{1.4}{\kelvin}$ in a magnetic field of \SI{0.75}{\tesla}. On-chip and off-chip coupling with \SI{> 35}{\percent} efficiency is achieved with a lensed fiber mounted on a piezoelectric nanopositioning device, as shown in Fig.~\ref{fig1:setup_and_inhom}b. The uncoupled end of the waveguide is terminated with a dielectric mirror of $\sim \SI{93}{\percent}$ reflectivity. The chip is probed using light from a frequency-stabilized laser system. The experimental setup is shown in Fig. \ref{fig1:setup_and_inhom}c and further described in the Methods section. To generate both strong optical pulses for optical pumping to prepare the quantum memory and faint pulses to be stored in the memory, we use acousto-optic (AOM) and electro-optic modulators (EOM). The retrieved light is detected by a superconducting nanowire single-photon detector after spectral and/or temporal filtering, which prevents detector blinding during memory preparation.

For initial device characterization, we use pulsed resonant fluorescence measurements~\cite{gritsch_narrow_2022}, as shown Fig.~\ref{fig1:setup_and_inhom}d. We find a clear fluorescence maximum at the emission frequency site A~\cite{gritsch_narrow_2022, rinner_erbium_2023, holzapfel_characterization_2025} with a full-width-half-maximum of \SI{352 \pm 15}{\mega \hertz}. This is fivefold narrower than previously measured in commercial samples~\cite{rinner_erbium_2023}, which we attribute to the reduced erbium concentration (see Methods).

\section{Spectral hole burning}\label{sec3}

\begin{figure}[htb]
\centering
\includegraphics[width=0.5\textwidth]{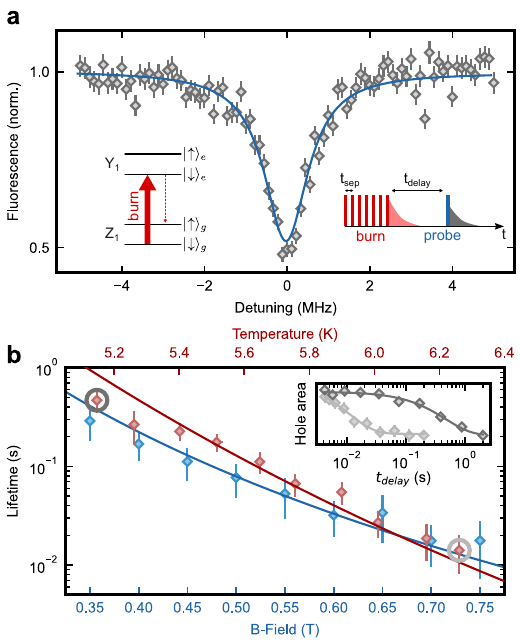}
\caption{\textbf{Spectral hole burning in Er:Si waveguides. a,} Left inset: Level scheme. Burn pulses are applied repeatedly on the $\ket{\downarrow}_\text{g} \rightarrow \ket{\downarrow}_\text{e}$ transition (thick red arrow). Spontaneous decay on the spin-flip transition (thin red arrow) transfers resonant dopants to $\ket{\uparrow}_\text{g}$. Right inset: Pulse sequence. \num{20} excitation pulses (red) of \SI{6}{\micro\second} duration each, separated by $t_\text{sep}=\SI{100}{\micro\second}$ are applied at the center of the inhomogeneous line. After spontaneous decay (light red) during a delay of $t_\text{delay}=\SI{1.5}{\milli\second}$, a probe pulse (blue) is applied, and the fluorescence (grey) is recorded within a \SI{300}{\micro\second} interval, and averaged over \num{4e3} repetitions (main panel: grey data). A Lorentzian fit (blue curve) yields a FWHM of \SI{1.29 \pm 0.10}{\mega \hertz} and an amplitude of \num{-0.48 \pm 0.02}.
\textbf{b,} Inset: The decay of the spectral holes as a function of $t_\text{delay}$ depends on the magnetic field and temperature. At \SI{20}{\milli\tesla} and \SI{6.5}{\kelvin} (light grey data), an exponential fit (solid) gives a spectral hole lifetime of \SI{11.6 \pm 3.8}{\milli\second}, while \SI{468.7 \pm 104.7}{\milli\second} are obtained at \SI{20}{\milli\tesla} and \SI{5.15}{\kelvin} (dark grey). Main panel: Spectral hole lifetimes as a function of magnetic field (blue data, bottom axis) at \SI{1.3}{\kelvin} and at different temperatures (red data, top axis) at \SI{20}{\milli\tesla}. Fits (solid lines) to phonon-induced relaxation processes allow extracting the coefficients of the Orbach (red) and direct (blue) decay processes. Error bars: 1 S.D. in all panels} \label{fig2:hole_burning}
\end{figure}

The AFC memory protocol~\cite{afzelius_quantum_2015} is based on the preparation of a spectrally periodic absorption pattern, created by spectral hole burning within the ensemble's inhomogeneous broadening. To this end, we apply 20 "burn" laser pulses, each with a pulse duration of \SI{6}{\micro\second}, a power of $\sim\SI{0.5}{\micro\watt}$ within the waveguide, and a pulse separation of \SI{100}{\micro\second} (i.e. approximately the excited state lifetime), as shown in the right inset of Fig.~\ref{fig2:hole_burning}a. Via spontaneous decay on the spin-flip transition (see level scheme in the left inset), this optical pumping sequence transfers a sub-ensemble of the dopants that is resonant with the pulses into a different spin state. Owing to the applied magnetic field, the optical transitions of this sub-ensemble occur at a different frequency.

After a controlled delay, the system is excited again with a single probe laser pulse of \SI{6}{\micro \second} duration and \SI{1.44}{\micro \watt} power, shifted in frequency using AOMs. The resulting fluorescence, measured within \SI{300}{\micro\second} after the pulse, reveals spectrally-selective population redistribution as a dip in the signal, called a spectral hole, as shown in \ref{fig2:hole_burning}a. The choice of the number, power, and pulse length of the burn pulses is optimized to minimize power broadening and heating, while maintaining high population transfer via optical pumping. Switching the burn laser off between pulses minimizes hole-broadening by laser-induced spectral diffusion~\cite{fruh_spectral_2026}. The probe pulse power and duration are chosen to optimize the signal-to-noise ratio. Its frequency is changed between repetitions to avoid hole-burning effects from the probe laser.

After spectral hole burning, the redistributed spin population decays back to the ground state via spin-lattice relaxation \cite{wolfowicz_quantum_2021}, leading to an exponential decay of the spectral hole as shown in the inset of Fig.~\ref{fig2:hole_burning}b. Measuring the hole lifetime as a function of temperature and magnetic field (main panel) allows the extraction of the spin-lattice relaxation parameters of Er:Si in site A. To obtain a higher signal, a chip with a higher Er concentration has been used in these measurements (see Methods); still, at the used parameters, interactions between the dopants do not contribute significantly to the spectral-hole decay ~\cite{car_optical_2019}.

The dominant spin-lattice relaxation mechanism depends on the temperature and magnetic field. At low field, two-phonon processes will dominate~\cite{wolfowicz_quantum_2021}. At elevated temperature, we find good agreement with an exponential temperature dependence ($T_1 = \alpha_\text{Orb}^{-1} \left(E_\text{CF}/k_B\right)^{-3}\text{e}^{E_\text{CF}/ (k_B T)}$), characteristic of an Orbach process (red fit curve) coupling to the closest crystal field level ($E_\text{CF}/h=\SI{2.6342}{\tera\hertz}$) with a coefficient of $\alpha_\text{Orb}=\SI{27 \pm 4e3}{\second^{-1}\kelvin^{-3}}$. At lower temperatures, the Orbach contribution to spin relaxation becomes negligible, and the spin lifetime exceeds many seconds, making direct measurements in our current devices impractical. However, upon application of a magnetic field, the direct phonon relaxation will be increased~\cite{shrivastava_theory_1983} $T_1 = \alpha_\text{dir}^{-1} g_\text{eff}^{-3} B^{-5} \coth{[\mu_B g_\text{eff}B/(k_B T)]}$, where $g_\text{eff}$ is the effective g-factor. The magnetic-field dependence follows this scaling, and a fit (blue solid line) gives the coefficient of the direct process as $\alpha_\text{dir}=\SI{0.23 \pm 0.03}{\second^{-1}\tesla^{-5}}$.

\section{AFC memory}\label{sec4}

\begin{figure}[htb]
\centering
\includegraphics[width=0.5\textwidth]{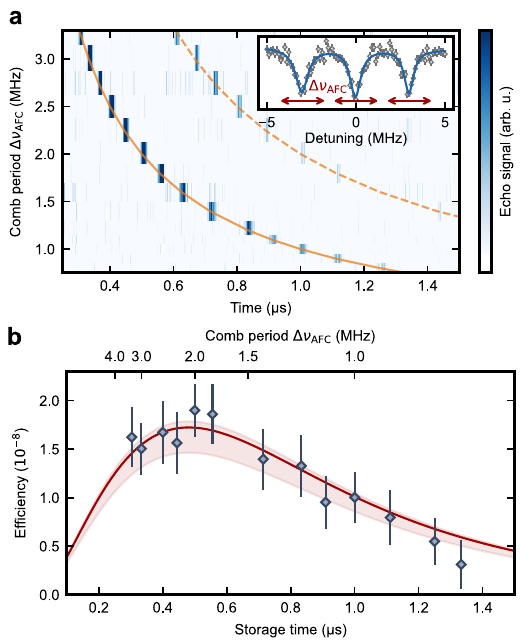}
\caption{\textbf{Quantum memory in an Er:Si waveguide. a,} Inset: Central part of the prepared AFC. After irradiating 20 burn laser pulses, frequency-modulated to form a regular comb with a period of \SI{3}{\mega \hertz}, the normalized fluorescence after a probe pulse reveals a periodic pattern of spectral holes. Lorentzian fits (blue) reveal a hole width of \SI{0.74 \pm 0.07}{\mega\hertz}. Main panel: Quantum memory experiment. Retrieved light after storing a faint, \SI{9.9\pm0.2}{\nano \second} long laser pulse in the erbium ensemble contained in the nanophotonic waveguide, averaged over \num{7.5e6} measurements. The storage time is pre-programmed by the comb period $\Delta\nu_\text{AFC}$, such that the light is retrieved after $1/\Delta\nu_\text{AFC}$ (orange); a weaker second echo is obtained at $2/\Delta\nu_\text{AFC}$ (orange dashed).
\textbf{b,} Memory efficiency. The retrieved light is compared to the input to determine the end-to-end efficiency. The efficiency drops at longer storage times (bottom axis) as the comb period (top axis) approaches the spectral hole width. A fit (red line) to the expected efficiency (eq.~\ref{eq:eff}) exhibits an optical depth of \SI{2.6 \pm 0.1e-3}{} (see main text). 
Error bars: 1 S.D. in all panels.
}\label{fig3:comb_and_memory}
\end{figure}

The storage of light pulses based on the AFC protocol uses spectral hole burning within the inhomogeneous line. Specifically, the absorption in the waveguide is tailored such that it consists of spectrally periodic features with a separation of $\Delta\nu_\text{AFC}$ \cite{afzelius_multimode_2009}. This comb period determines the storage time $t_\text{s}=1/\Delta\nu_\text{AFC}$. In turn, the overall width of the comb, given by the number of comb teeth $N$ times $\Delta\nu_\text{AFC}$, determines the usable bandwidth of the memory and thus the minimal duration $\tau_\text{p,min}$ of an impinging light pulse to be stored: $\tau_\text{p,min} \gtrsim (N\cdot \Delta\nu_\text{AFC})^{-1}$. This holds only as long as the AFC does not exceed the inhomogeneous broadening; based on the results in Fig.~\ref{fig1:setup_and_inhom}, this requires a minimum pulse duration of $\gtrsim\SI{3}{\nano\second}$.

In our experiment, we generate a comb pattern with $N=36$ lines. To this end, sidebands are modulated to the frequency-stabilized burn laser using two EOMs with modulation frequencies $\Delta\nu_\mathrm{EOM1}=12 \Delta\nu_\text{AFC}$ and $\Delta\nu_\mathrm{EOM2}=4\Delta\nu_\text{AFC}$. The sideband power is adjusted to produce nine lines with almost equal intensity and a spacing of $4\Delta\nu_\text{AFC}$. Using AOMs, this pattern is subsequently applied four times, each time with an additional frequency shift of $\Delta\nu_\text{AFC}$. The entire sequence is then repeated 5 times, yielding a total of 20 sequential burn pulses. The pulse parameters for the burn pulses are the same as in Fig.~\ref{fig2:hole_burning}a. Thus, the EOM modulation reduces the burn power applied at each frequency approximately ninefold. The inset of Fig~\ref{fig3:comb_and_memory}a shows the central structure of the resulting AFC, measured using pulsed fluorescence spectroscopy. A Lorentzian fit of the comb lines gives \SI{743 \pm 72}{\kilo\hertz} FWHM and a relative decrease of the fluorescence by \SI{26 \pm 1}{\percent}. 

In the following, we use the generated AFC in a quantum memory experiment. To this end, faint laser pulses of $\tau_p=\SI{9.9\pm0.2}{\nano \second}$ duration that contain $\SI{21 \pm 2e3}{}$ photons on average are coupled into the waveguide. After being absorbed by the AFC, the light is reemitted after a time $t=1/\Delta\nu_\text{AFC}$ that can be programmed by the comb period. This storage experiment is repeated 25 times with a delay of $\SI{1.6}{\micro \second}$, i.e., exceeding the storage time, before the comb preparation starts over. Fig.~\ref{fig3:comb_and_memory}a (main panel) shows the signal obtained when varying the comb period, averaged over \SI{7.5e6}{} repetitions. Readout photons are obtained at the expected emission time (solid orange line) and at the second-order echo (at $2/\Delta\nu_\text{AFC}$, orange dashed), which originates from the spectral shape of the prepared comb absorption feature~\cite{afzelius_multimode_2009}. 

To determine the memory efficiency, the recorded fluorescence counts within $\pm \SI{30}{\nano \second}$ around the storage time are summed and compared with the intensity of the input pulses (see Methods). The efficiency peaks at a storage time of \SI{0.5}{\micro\second} at a value of \SI{1.9 \pm 0.28e-8}{} as shown in Fig.~\ref{fig3:comb_and_memory}b. From the measured AFC absorption feature, we calculate the expected memory efficiency (see methods) and fit the data with the optical depth of \SI{2.6 \pm 0.1e-3}{} being the only fit parameter. Our analysis shows that, for a fixed optical depth, the optimal storage time is inversely proportional to the hole width in the AFC absorption spectrum --- the narrower the holes, the longer the storage time. In our current experiment, the latter is thus limited by laser-induced spectral diffusion~\cite{fruh_spectral_2026}, which broadens the holes compared to the homogeneous linewidth of $\SI{10}{\kilo\hertz}$~\cite{gritsch_narrow_2022}. Optimized materials and comb preparation sequences may thus increase the achieved value to tens of microseconds. Achieving even longer memory times requires transferring the excited-state population to long-lived electronic or nuclear spin states.

\begin{figure}[htb]
\centering
\includegraphics[width=0.5\textwidth]{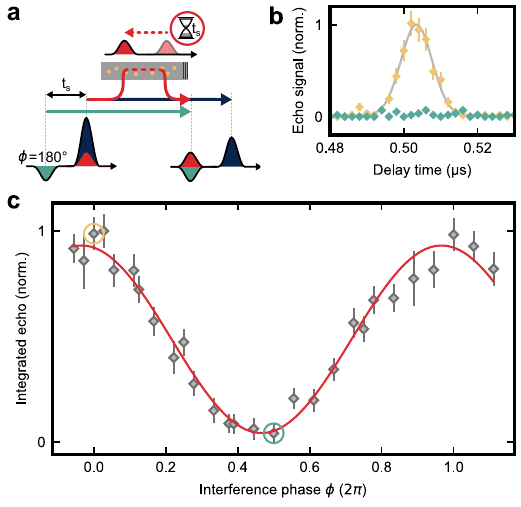}
\caption{\textbf{Coherence measurement. a,} Pulse sequence. The memory is probed by sending in faint laser pulses (bottom left, red and dark blue). To demonstrate that the fraction of the pulse (red) that is stored and retrieved from the memory (top) is phase coherent with the input light, we interfere it with a second laser pulse (green) with the same envelope as the storage pulse, sent with a matching delay $t_s$ and a controlled phase $\phi$. \textbf{b,} Phase-dependent readout signal. We program the AFC to a storage time of \SI{0.5}{\micro \second} and irradiate \SI{9.9\pm0.2}{\nano \second} long Gaussian pulses; the sequence is averaged over \SI{5e6}{} repetitions. Yellow: When the interference pulse has the same phase as the input, constructive interference leads to the observation of a strong signal at the programmed delay. Green: When the phase is changed by $\pi$, destructive interference leads to an almost complete disappearance of the signal. \textbf{c,} Continuous phase sweep. Fitting a sinusoidal oscillation (red) to the interference signal integrated over the pulse duration (gray data) gives a visibility of \SI{91.3 \pm 3.0}{\percent}. The green and yellow circles mark the measurements in panel b. Error bars: 1 S.D. in all panels.}
\label{fig4:coherence}
\end{figure}

After determining the storage time, we investigate the coherence of the retrieved light. To this end, we interfere the readout signal with a second laser pulse that exhibits the same temporal shape as the impinging storage pulse. The time and amplitude at which the second pulse is sent are chosen such that a high interference contrast is obtained, as schematically shown in Fig.~\ref{fig4:coherence}a. The phase of the second pulse relative to the storage pulse is controlled via the AOMs. When the readout light is in phase with the interference pulse (yellow data in Fig.~\ref{fig4:coherence}b), constructive interference leads to a large readout signal that almost vanishes completely when the readout light is out of phase (green). Again, the experiment is repeated 25 times before the comb preparation starts over, and the signal is averaged over \SI{5e6}{} repetitions in total. When integrating the fluorescence over the entire readout pulse duration, a sinusoidal interference is observed as a function of the phase of the interference laser pulse, as shown in Fig.~\ref{fig4:coherence}c. From a fit to the data, we extract a visibility of \SI{91.3 \pm 3.0}{\percent}, demonstrating the coherence of the memory. The reduction from perfect interference may be due to the laser's frequency instability (with a locked linewidth of $\SI{39 \pm 1}{\kilo\hertz}$), optical path instability, and imperfect matching of the interference pulse envelope. The given value can therefore be considered a lower bound for the device's performance.

The large overlap between the memory and interference pulses allows us to precisely estimate the memory's efficiency by calibrating the amount of attenuation required to optimize the overlap. We find an attenuation of \SI{1.89 \pm 0.03e-8}{} when using the same parameters as for the measurements in Fig.~\ref{fig3:comb_and_memory}, and use this as a calibration parameter for the efficiencies. Note that the above efficiency does not include the fiber-to-chip coupling efficiency ($\eta_\text{fc}=\SI{38 \pm 3}{\percent}$), which may be improved in future devices and is not relevant for on-chip applications.

\section{Discussion}
Our measurements demonstrate the phase-coherent storage of faint laser pulses in an on-chip silicon nanophotonic quantum memory. At a storage time of \SI{1}{\micro\second}, the achieved efficiency outperforms that of silicon transmission lines by almost \num{400} orders of magnitude. However, the efficiency is not yet high enough to test the device with single photons or to use it in practical applications. To determine the main limitation, we calculate the efficiency $\eta$ expected for a periodic AFC absorption pattern with a Finesse $F > 1$ and an atomic ensemble with optical depth  $d=F\tilde{d}$~\cite{bonarota_efficiency_2010,zang_provable_2025}. In the case of our experiment, the absorption pattern features Lorentzian holes with a relative hole depth $A$; for this, we find (see methods): 
\begin{equation}
\eta=\left(\frac{A\tilde{d}}{2}\right)^2 e^{-\tilde{d}\left(F-A \arctan{\left[F\right]} \right)} \left(c_{+} e^{\pi/F} + c_{-} e^{-\pi/F} \right)^2,\label{eq:eff}
\end{equation}
with the coefficients 
\begin{equation}
    c_{+}=\pi +  \Im\left\{{\mathrm{E_1}\left[-\frac{\pi}{F}-i\pi\right]}\right\},\ c_{-}=\Im\left\{\mathrm{E_1}\left[\frac{\pi}{F}+i\pi\right]\right\},
    \label{eq:coeff}
\end{equation}
where $\mathrm{E_1}$ is the exponential integral function~\cite{abramowitz_handbook_1965}.
At low optical depth, the efficiency increases with the square of the effective hole depth, and an optimized burn scheme achieving full contrast would lead to an increase of a factor of \num{15} for a storage time of \SI{0.5}{\micro\second} using the same spectral properties of the erbium ensemble.

Further improvement would require an increased optical depth and switching to a Lorentzian peak absorption structure. The latter would reduce the effect of background absorption, yielding an optimal optical depth of $d=2F/arctan(F)\approx4.4$ for a fixed Finesse of $F\approx2.7$. With this fixed Finesse set by the spectral hole width, one would expect an efficiency of $\eta\approx 0.1$. The required optical depth is \num{1.7e3} times the current value.
Different strategies can be followed to achieve this improvement: First, the waveguide length can be increased. The loss of our foundry-made waveguides~\cite{aalto_open-access_2019} is around \SI{0.1}{\decibel\per\centi\meter}, which limits practical lengths to $\lesssim\SI{30}{\centi\meter}$ before the efficiency drops by $\SI{3}{\decibel}$. This would already increase the optical depth by a factor of \num{25} compared to the waveguides used in this work. However, this would come at the price of a corresponding increase in the device footprint. Second, the integration of dopants into slow-light waveguides~\cite{burger_inhibited_2025} can strongly enhance the absorption cross section of individual emitters. Finally, the optical depth can be further increased by using a higher dopant density, potentially combined with a more homogeneous spread across the waveguide mode that can be achieved using multiple implantation energies.

On thin-SOI waveguides fabricated in an academic cleanroom, \num{200}-fold larger erbium peak concentrations with a ten-fold larger spread in the implantation direction have been studied, without a significant increase in the inhomogeneous linewidth~\cite{gritsch_narrow_2022}. However, the devices had a higher loss of $\SI{5.7 \pm 1.2}{\decibel \per \centi\meter}$, hindering the simultaneous use of longer waveguides described earlier. In addition, laser-induced spectral diffusion observed at higher concentrations~\cite{fruh_spectral_2026} may impede the maximum storage time in such devices. Thus, increasing the dopant density without being limited by the mentioned effects will require a systematic optimization of the sample fabrication, such that the fraction of erbium dopants that is integrated into site A is enhanced from its current value of $\lesssim \SI{1}{\percent}$. 

Combining these approaches, it seems that optical depths for optimal forward read-out efficiencies at the current spectral properties of the ensemble are attainable. Further efficiency increases require backward read-out or cavity-enhanced memories~\cite{afzelius_multimode_2009}. The latter approach seems particularly promising as it simultaneously overcomes the forward read-out limitations and strongly enhances the absorption cross-section of each dopant~\cite{afzelius_quantum_2015, zhong_nanophotonic_2017, craiciu_multifunctional_2021, reiserer_colloquium_2022, meng_efficient_2026}. By increasing the device's efficiency towards the fundamental limits of the memory protocols, applications in photonic quantum computing come within reach. In particular, with active photonic components~\cite{aharonovich_programmable_2026}, such quantum memory would enable synchronisation in qubit-based architectures~\cite{rudolph_why_2017, slussarenko_photonic_2019, alexander_manufacturable_2025}, and replace fiber-optical delay lines in temporally-multiplexed hybrid architectures ~\cite{takeda_toward_2019, maring_versatile_2024}, paving the way towards scalability.

Combining these capabilities with on-demand readout and longer storage times would furthermore enable a wide variety of applications in distributed quantum information processing. To this end, electrodes could be used to induce controlled shifts in the optical transition frequencies via the Stark effect \cite{liu_-demand_2022, craiciu_multifunctional_2021}. Alternatively, the excitation could be transferred into a spin-wave excitation \cite{afzelius_demonstration_2010} of the hyperfine levels of the \ce{^167Er} isotope, which have achieved second-long coherence in other materials \cite{rancic_coherence_2018}. Thus, by implementing suitable long-term memory protocols~\cite{zhou_photonic_2023} and leveraging the earlier-discussed efficiency increase, we expect that the foundry-made quantum memories demonstrated in this work may become a key component not only in quantum computing systems but also in future long-distance quantum networks and quantum repeaters.

\section{Methods}\label{sec:methods}

\subsection{Sample fabrication}

The silicon photonic chip used for the measurements presented in Fig.~\ref{fig1:setup_and_inhom},\ref{fig3:comb_and_memory},\ref{fig4:coherence}, and Fig.~\ref{fig2:hole_burning}a of this work was commercially fabricated by VTT in Finland, as described in the main text. Erbium was implanted with mixed isotopic abundance by the Helmholtz Zentrum Dresden Rossendorf (HZDR) using a dose of \SI{5e10}{\per \centi \meter \squared} and an energy of \SI{3}{\mega \electronvolt}. According to simulations (SRIM)~\cite{ziegler_srim_2008}, this results in an approximately Gaussian implantation profile with a peak concentration of \SI{1e15}{\per \centi\meter^3} at a depth of \SI{1}{\micro\meter}, with a longitudinal straggle of \SI{0.1}{\micro\meter}. After implantation, rapid thermal annealing is performed at \SI{500}{\celsius} for \SI{5}{\min} to cure the crystal damage caused by implantation. 

The silicon photonic chip used for spin lifetime measurements in Fig. \ref{fig2:hole_burning}b was commercially obtained from Advanced Micro Foundry Pte Ltd. Measurements were performed on a ridge waveguides with \SI{220}{\nano \meter} height, \SI{500}{\nano \meter} width, and \SI{6.25}{\milli \meter} length, terminated by a mirror structure as described in~\cite{rinner_erbium_2023}. The sample was implanted by Ion Beam Services S.A.S. with a dose of \SI{1e12}{\per \centi \meter \squared} and an energy of \SI{250}{\kilo\electronvolt}, which gives a peak concentration of \SI{1e17}{\per \centi\meter^3}. The chip was annealed at \SI{470}{\celsius} for \SI{15}{\second} and \SI{500}{\celsius} for \SI{5}{\second}.

\subsection{Setup}

The samples are cooled to $\lesssim \SI{1.4}{\kelvin}$ in a closed-cycle vacuum cryostat (DRY ICE 1K from ICEoxford Ltd.). A superconducting solenoid magnet (Cryomagnetics Inc.) provides fields of up to \SI{3}{\tesla}. Using silver conductive adhesive, the samples are mechanically and thermally anchored to a printed circuit board screwed onto the cold stage. For on- and off-chip coupling via a 95:5 beam splitter (Evanescent Optics 954), a single-mode lensed fiber (AMS Technologies TSMJ) is aligned with the sample using a cryogenic piezo positioner stack (attocube ANPx311, ANPz102).

The laser pulses are generated from a tunable diode laser (Toptica CTL) stabilized to a frequency comb (Toptica DFC). To this end, we use two \SI{300}{\mega \hertz} AOMs (Gooch and Housego PM Fiber-Q 1550) for high-contrast switching (with rise- and fall-times of $\sim\SI{10}{\ns}$), two phase EOMs (iXblue/Exail MPZ-LN-10) for generating sidebands, and one IQ EOM (iXblue MXIQER-LN-30) for pulse shaping on \SI{}{\nano \second} timescale. These devices are controlled by an arbitrary waveform generator (Zurich Instruments HDAWG). Light is detected by a superconducting nanowire single photon detector (PhotonSpot). To prevent detector blinding during resonant excitation, we use a fast optical switch (Agiltron UltraFast Dual Stage $1\times1$). A fiber-based polarizer (Thorlabs), and fiber polarization controllers (Thorlabs) control the input polarization. A fiber-optical isolator (Newport ISC-1550) is used to block multiple pulse reflections in the setup. In addition, the memory experiments in Fig. \ref{fig3:comb_and_memory} and Fig. \ref{fig4:coherence} use a bandpass filter with a \SI{100}{\pico \meter} transmission window (WL Photonics WLTF-NE-S-1550) to block the fluorescence of off-resonantly excited erbium dopants~\cite{gritsch_narrow_2022}.

\subsection{Spin lifetime measurements}

The hole-burning scheme described in section~\ref{sec3} is used to measure the dynamics of spin population redistribution toward equilibrium. The spectral hole area is proportional to the amount of population pumped away from thermal equilibrium. Measuring the dynamics of the hole area allows us to extract the spin relaxation dynamics. We extract the hole area by numerically integrating the fluorescence probe signal versus frequency using the trapezoidal rule, after subtracting an offset determined by the average of the \num{40} points with the largest detected signal within the same scan. The spectral hole area is probed at logarithmically spaced delay times with respect to the burn pulse. 

At cryogenic temperatures and in low magnetic fields, the spin lifetimes get very long. This makes sequential acquisition of all 10 data points in the inset of Fig.~\ref{fig2:hole_burning}b impractical. Therefore, instead of measuring the spectral holes at each burn-probe delay individually, we use ten probe pulses after each hole preparation. To prevent the extracted decay constant from being affected by repeated probing at the same frequency, the probe frequency is varied between pulses. Thus, only one part of the hole is measured for each delay, and the order is changed between repetitions until the full spectra can be combined and fitted. The extracted lifetime from the exponential fit yields the decay curves shown in Fig.~\ref{fig2:hole_burning}b.

The temperature-dependent data of Fig.~\ref{fig2:hole_burning}b are fitted using a simplified function for the Orbach two-phonon relaxation. We extract the Orbach coefficient by fitting logarithmically rescaled lifetime data with a linear model with a fixed slope equal to the crystal-field splitting to the closest crystal field level. The uncertainty of the extracted Orbach coefficient is determined by the fitting error and the uncertainty in the crystal-field splitting.
In the lifetime data versus magnetic field strength, we fit the non-linear model for direct one-phonon relaxation to the data. The uncertainty of the extracted direct process coefficient is estimated using the fitting error, as well as uncertainties in the magnetic field, effective g-factor and temperature.

\subsection{Analysis of the memory efficiency}
The expression for the expected memory efficiency in eq.~\ref{eq:eff} is derived using a model of linearized Maxwell-Bloch equations~\cite{afzelius_multimode_2009} generalized to a periodic spectral absorption function as outlined in refs.~\cite{bonarota_efficiency_2010,zang_provable_2025}. This formula uses the rotating-wave and slowly-varying-envelope approximations and is valid for small optical depths, sufficiently short waveguide lengths relative to the stored pulse length, and negligible dispersion. Furthermore, it assumes that the effects of the homogeneous broadening of individual dopants, the finite bandwidth due to the inhomogeneous broadening of the ensemble, and the imperfections of the finite comb of Lorentzian holes in the ensemble's spectral absorption feature can be neglected. These approximations are satisfied for the parameters used in our memory protocol, with an inhomogeneous linewidth of \SI{351 \pm 16}{\mega\hertz}, a hole linewidth of \SI{0.74\pm0.07}{\mega\hertz}, a total of \num{36} holes, an optical depth of \SI{2.6 \pm 0.1e-3}{}, a waveguide length of $L=\SI{6.25}{\milli\meter}$, and a memory pulse of Gaussian shape with a temporal width of $\tau=\SI{9.9\pm0.2}{\nano\second}$. With these parameters, the spectral absorption feature of our AFC is approximated by the periodic function
\begin{equation}
    \frac{\alpha(\omega)}{\alpha_0}= \left( 1-A\sum_{n\in \mathbb{Z}} \frac{\left(\gamma/2\right)^2}{\left(\gamma/2\right)^2+\left( \omega/(2\pi)- n/t_\text{s}\right)^2} \right),
\end{equation}
with $A$ the relative hole depth, $\gamma$ the hole width, $\alpha_0$ the maximum absorption, and $t_\text{s}=1/\Delta\nu_\mathrm{AFC}$ the periodicity. The memory efficiency $\eta$ is then calculated using the Fourier series expansion of the periodic spectral absorption~\cite{zang_provable_2025}
\begin{equation}
    \eta=|F_{-1} \tilde{L}|^2 e^{-F_0\tilde{L}},
    \label{eq:eff_general}
\end{equation}
with $\tilde{L}$ the effective absorption length, and the Fourier coefficients
\begin{align}
    F_{0}&=\frac{t_\text{s}}{2\pi}\int_{-\pi/t_\text{s}}^{\pi/t_\text{s}}\alpha(\omega ) \mathrm{d}\omega,\\
    F_{-1}&=\frac{t_\text{s}}{2\pi}\int_{-\pi/t_\text{s}}^{\pi/t_\text{s}}\alpha(\omega ) e^{i\omega t_\text{s}}\mathrm{d}\omega.
\end{align}
With the Finesse $F=(\gamma t_\text{s})^{-1}$, the zeroth-order Fourier coefficient can be evaluated to
\begin{equation}
    F_0 = \alpha_0\left(1-\frac{A}{F} \arctan{(F)}\right),
    \label{eq:F0}
\end{equation}
and the first-order Fourier coefficient to
\begin{equation}
    F_{-1}= \frac{A\alpha_0}{2F} \left(c_+ e^{\pi/F} + c_- e^{-\pi/F}\right).
    \label{eq:F-1}
\end{equation}
Here, the coefficients are defined as in eq.~\ref{eq:coeff}, and we have used the symmetry property of the exponential integral function for complex conjugation~\cite{abramowitz_handbook_1965}
\begin{equation}
    \mathrm{E}_1\left[ \bar{z} \right] = \overline{E_1[z]}.
\end{equation}
Inserting the results of eq.~\ref{eq:F0} and eq.~\ref{eq:F-1} into the efficiency formula eq.~\ref{eq:eff_general}, we arrive at the expression of eq.~\ref{eq:eff}. This function is then used to perform a least-squares fit of the efficiency measurements, with the optical depth as the only free parameter. To estimate the uncertainty of the fit result, we use a resampling method. We repeatedly draw random parameter sets and data points from a normal distribution defined by the respective uncertainties and mean values, and apply our fitting routine. The standard deviation of the resulting estimates of the optical depth from resampling, repeated \num{5e4} times, is used to estimate the uncertainty in the optical depth extraction.

\section*{Funding}
\begin{itemize}
\item Deutsche Forschungsgemeinschaft (DFG, German Research Foundation) under the German Universities Excellence Initiative, GA EXC-2111 - 390814868
\item Munich Quantum Valley, which is supported by the Bavarian state government with funds from the Hightech Agenda Bayern Plus
\item German Federal Ministry of Research, Technology and Space (BMFTR), GA 13N16921 and 16KIS2198
\item European Union (ERC project OpENSpinS, GA 101170219). Views and opinions expressed are those of the authors only and do not necessarily reflect those of the European Union or the European Research Council. Neither the European Union nor the granting authority can be held responsible for them.
\end{itemize}

\end{document}